\documentstyle[12pt]{article}

\newcommand {\parn}{\par \noindent}
\newcommand{\num} {\hspace {-7 mm}\leqno}
\newtheorem{th}{\bf {Theorem }}
\newtheorem{lem}[th]{\bf {Lemma}}
\newtheorem{prop}[th]{\bf {Proposition}}
\newtheorem{cor}[th]{\bf {Corollary}}

\newcommand\n{I\hskip-4pt N}

\newcommand\r{I\hskip-3pt R}

\newcommand{\fta}{\Phi_{h , \omega}}
\newcommand{\ftb}{\Psi_{h , \omega}}

\newcommand{\ex}  {e^{ {i \over h} \sqrt{\lambda} x. \omega}}

\newcommand{\co} {\chi_0 ( h^{\epsilon} D)}

\begin{document}

\centerline{\large {\bf{An inverse scattering problem for the Schr\"
odinger}}} \centerline{\large {\bf{equation in a semiclassical
process.}}}

\vspace {3 cm} \noindent \centerline{\small{\bf{Fran\c{c}ois
Nicoleau}}}\parn \centerline{\vspace {5 mm}}
\centerline{\small{D{\'e}partement de Math{\'e}matiques}}\parn

\centerline{\small{U.M.R 6629 - Universit{\'e} de Nantes}}\parn
\centerline{\small{2, rue de la Houssini{\`e}re  BP 92208 }}\parn
\centerline{\small{F-44322 Nantes cedex 03}}\vspace {0,1cm}\parn
\centerline{\small{e-mail : nicoleau@math.univ-nantes.fr}}\parn

\vspace{3 cm} \parn
\begin{abstract}
We study an inverse scattering problem for a pair of Hamiltonians
$(H(h) , H_0 (h))$ on $L^2 (\r^n )$,  where $H_0 (h) = -h^2
\Delta$ and $H (h)=  H_0 (h) +V$, $V$ is a  short-range potential
with a regular behaviour at infinity and $h$ is the semiclassical
parameter. We show that, in dimension $n \geq 3$,  the knowledge
of the scattering operators $S(h)$, $h \in ]0, 1]$, up to
$O(h^\infty)$ in ${\cal{B}} (L^2(\r^n ))$,  and which are localized
near a fixed energy $\lambda >0$, determine the potential $V$ at
infinity.
\end{abstract}

\vspace{1 cm}
\section{\bf{Introduction.}}
\par
This work is a continuation of [10]. In the present paper, we study  an inverse scattering problem for the pair
of Hamiltonians
$(H(h),H_0 (h) )$ on $L^2 (\r^n)$, $n \geq 2$, where the free operator is
$H_0 (h) =  -h^2 \Delta, \ h \in ]0, 1 ]$ is the semiclassical parameter, and the perturbed Hamiltonian is given by :
$$
H (h)=   H_0 (h) +V . \num (1.1)
$$
We assume that $V$ is a real-valued smooth short-range potential satisfying :
$$
\forall \alpha \in \n^n \ \ , \ \ \exists \ C_{\alpha} >0 \ \ , \  \  \mid \partial_x^\alpha  V(x) \mid \leq \ C_{\alpha} \
<x>^{-\rho-\mid\alpha\mid} \ ,\ \rho>1. \num (H_1)
$$

\newpage

\parn
Under the hypothesis $(H_1)$, it is well-known that the wave operators :
$$
W^{\pm} (h) \ = \ s-\lim_{t \rightarrow \pm \infty} \ e^{it H(h)} e^{-it H_0 (h) } ,
\num (1.2)
$$
exist and are complete, i.e $Ran \ W^{\pm} (h)  = { \cal{H}}_{ac} (H)$ = subspace of absolute continuity of $H(h)$.

\vspace{0,5cm}
\parn
Let $S(h)$ be the scattering operator defined by :
$$
S(h) =  W^{+*} (h) \ W^{-} (h) . \num (1.3)
$$

\vspace{0,5cm}
\parn
Since $S(h)$ commutes with $H_0 (h)$, we can define the scattering matrices
$S(\lambda, h)$, $\lambda >0$,  as unitary operators acting on $L^2(S^{n-1})$, where  $S^{n-1}$ is the unit
sphere in $\r^n$. We denote by $\Phi_0 (x, \lambda, \omega,h), \ (\lambda, \omega) \in ]0, +\infty[
\times S^{n-1}$, the generalized eigenfunction of $H_0 (h)$ :
$$
\Phi_0 (x, \lambda, \omega,h) = e^{{i \over h} {\sqrt{\lambda}} x \cdot \omega}.
\num (1.4)
$$
The unitary mapping ${\cal F}_0 (h)$ defined by :
$$
({\cal F}_0 (h) f) (\lambda, \omega) = (2\pi h)^{-{n \over 2}} \  \lambda^{ {n-2} \over 4} \
\int_{\r^n} \ {\overline{\Phi_0}} (x, \lambda, \omega,h) \ f(x) \ dx \ ,
\num (1.5)
$$
gives the spectral representation for $H_0 (h)$, i.e $H_0 (h)$ is transformed into the multiplication by $\lambda$
in the space $L^2\ ( \r^+ \ ;\ L^2(S^{n-1}))$.

\vspace{0,5cm}
\parn
Then, the scattering matrices are defined by :
$$
{\cal F}_0 (h) S(h) f (\lambda) = S(\lambda, h) \ ({\cal F}_0 (h) f) (\lambda),
\num (1.6)
$$
and we have the Kuroda's representation formula :
$$
S(\lambda, h) = 1 - 2i\pi F_0 (\lambda, h) V F_0 (\lambda, h)^* +2i\pi F_0 (\lambda, h)V R(\lambda +i0,h) V
F_0 (\lambda, h)^* \ ,
\num (1.7)
$$
where for $f \in L_s^2(\r^n), \ s>{1 \over 2}$,
$F_0 (\lambda, h) f (\omega) = ({\cal F}_0 (h) f) (\lambda, \omega)$ and $R(\lambda +i0,h)$ is given by the well-known
principle of limiting absorption.

\vspace{1 cm}
\parn
We recall the non-trapping condition. Let ${\displaystyle{\left( z(t,x, \xi), \ \zeta(t,x,\xi) \right)}}$ be the
solution to the Hamilton equations :
$$
\dot{z} (t,x, \xi) = 2 \zeta (t,x, \xi) \ \ ,\ \
\dot{\zeta}(t,x, \xi) = - \nabla V(z (t,x, \xi)),
\num (1.8)
$$
with the initial data $z(0, x, \xi) =x, \ \zeta (0,x,\xi) = \xi$.
\vspace{0,5cm}\parn
We say that the energy $\lambda$ is non-trapping, if for any $R \gg 1$ large enough, there exists $T=T(R)$ such that
$\mid z(t,x,\xi)\mid >R$ for $\mid t\mid >T$, when $\mid x\mid <R$ and $\lambda = \xi^2+V(x)$.

\vspace{0,2cm}
\parn
When $\lambda$ is a non-trapping energy, and if $V$ satisfies $(H_1)$ with $\rho>0$, we have the following estimate
when $h \rightarrow 0$, (see [12], [14]),
$$
\forall s>{1 \over 2} \ \ ,\ \ \mid \mid <x>^{-s} \ R(\lambda+i0,h) \ <x>^{-s} \mid\mid = O\ ({1 \over h}).
\num (1.9)
$$

\vspace{1cm}
\parn
The goal of this paper is to give a partial answer to the
following conjecture for a class of potentials which are regular at infinity.
We denote by ${\cal B} ({\cal H})$ the space of bounded operators acting on a
Hilbert space ${\cal{H}}$, and by $f_+ (x) = max \ (f(x),0)$.

\vspace{0,5cm}

\parn
{\bf{Conjecture.}}

\vspace{0,2cm}\parn
Let $V_1$ and $V_2$ be potentials satisfying $(H_1)$ and let
$S_j (\lambda, h)$, $j=1,2$, be the corresponding scattering matrices at a fixed non-trapping energy $\lambda>0$.
Assume that, in the semiclassical regime  $h \rightarrow 0$, $S_1 (\lambda, h)= S_2 (\lambda, h) + O(h^{\infty})$,
in ${\cal B} (L^2(S^{n-1}))$.
\vspace{0,1cm}\parn
Then, the potentials are equal in the classical allowed region, i.e
$$
(\lambda-V_1 (x))_+ = (\lambda-V_2 (x))_+ .
\num (1.10)
$$

\vspace{0,8cm}
\parn
According to the author, this result is not known in the general case. In [11], for
potentials satisfying $(H_1)$ with $\rho>n$, Novikov shows, without the non-trapping condition and using the
${\overline{\partial}}$-method,
that if $S_1(\lambda, h) = S_2(\lambda, h) \ ,\ \forall h\in]0,1[$, then $V_1 =V_2$. It is clear that this approach is not adapted to
study our conjecture in the semiclassical setting since, physically, we can not obtain any information  on the
potential outside the classical allowed region. To give an  example, in [13], Robert and Tamura obtain,
under suitable assumptions, the complete asymptotic expansion of the scattering amplitude $f(\lambda, \theta, \omega, h)$,
with $\theta, \omega$ fixed, $\theta \not=\omega$, and where $\lambda$ is a fixed non-trapping energy.
The coefficients of this expansion depend only on the values of $V(x)$ in $\{ x \in \r^n \ : \ V(x) \leq \lambda\}.$
It follows that if we modify the potential $V$ outside the classical allowed region, the scattering amplitude remains
unchanged up to $O (h^{\infty})$.

\vspace{0,3cm}\parn
It seems to be difficult to prove our conjecture for a fixed energy without assuming the non-trapping condition.
The difficulty comes from the estimate of the resolvent $R(\lambda+i0,h)$ due to
resonances converging exponentially to the real axis. To avoid the problem
due to these resonances, one can average with respect to the  energy $\lambda$.
In [16], Yajima  obtains, without the non-trapping condition and
with the additional condition $\rho>max\ (1, { {n-1} \over 2})$, the asymptotic expansion of the scattering amplitude,
averaging with respect to $\lambda$ and $\theta$ . In [6], Michel proves for a similar result when one averages only with
respect to $\lambda$, i.e. $\theta \not= \omega$ are fixed. In [7], he obtains for a fixed trapping energy, but with
an additional assumption on the resonances, the asymptotic expansion of the scattering amplitude similar to the one
established in the non-trapping case.

\vspace{0,3cm}\parn
For potentials with compact support, an inverse scattering problem close to our conjecture
is studied in [1]. Assuming that ${\displaystyle{\lambda> \sup_{x\in \r^n} V_+ (x)}}$
and that the metric $g(x)= (\lambda-V(x))\  dx^2$ is
conformal to the Euclidian, Alexandrova shows in [1] that the scattering amplitude in the semiclassical regime,
(actually, the scattering relations), determines $V$ uniquely. This result is obtained by comparing the inverse
scattering problem for the Schr\" odinger equation with the problem of the wave equation with variable speed for which
the result is well-known. The scattering relations for the metric $g(x)$ determine the boundary distance function uniquely,
and then we determine $g$ and therefore $V(x)$. This method can not be used for general potentials satisfying $(H_1)$.

\newpage
\parn
Now, let us explain more precisely the setting of this paper. We consider the class of potentials $V$ which are asymptotic
sums of homogenous terms at infinity, i.e. we assume that $V \in C^{\infty} (\r^n )$ and satisfies the regular
behaviour at infinity :
$$
V(x) \ \simeq \ \sum_{j=1}^{\infty} \ V_j (x)
\num (H_2)
$$
where the $V_j (x)$ are homogeneous functions of order $-\rho_j$ with $1<\rho_1<\rho_2< \cdots$ .

\vspace{0,3cm}\parn
Inverse scattering at a fixed energy with regular potentials at infinity and when the Planck's constant $h=1$ was first
studied in [5] with $\rho_j =j+1$. In [15], Weder and Yafaev study  a similar problem for general degrees
of homogeneity $\rho_j$. They show that the complete asymptotic expansion of the potential is determined by the singularities in the
forward direction of the scattering amplitude. In particular, they obtain the following result :
\vspace{0,3cm}\parn
Let $V_1, \ V_2$ be potentials satisfy the assumption $(H_2)$ and let $S_j(\lambda)$, $j=1,2$,
be the corresponding scattering matrices. Assume that the kernel of $S_1 (\lambda)-S_2(\lambda)$ belongs to
$C^{\infty} (S^{n-1} \times S^{n-1})$, then $V_1 -V_2$ belongs to the Schwartz class ${\cal S} (\r^n)$.

\vspace{0,3cm}\parn
If we want to study the inverse scattering problem in the semiclassical setting with a fixed non-trapping energy and
for potentials having a regular behaviour at infinity, it seems to the author that the technics developed in [15]
are not well adapted. Indeed, as it was pointed in ([12], [13]), for potentials satisfying $(H_1)$ with $\rho>n$,
the forward amplitude $f(\lambda, \omega, \omega, h)$ is of order $O\ (h^{-\mu})$ with
${\displaystyle{\mu= { {(n-1)(n+1)} \over {2(\rho -1)}}}}$ and
$f(\lambda, \theta, \omega, h)$ with $\theta\not=\omega$ is of order $O (1)$. Thus, the scattering amplitude has a strong
peak in a neighborhood of $\omega$.
\vspace{0,2 cm}\parn
Moreover, let us emphasize that, generically,  we do not have any information on the kernel of $S_1 (\lambda, h)- S_2 (\lambda, h)
= O(h^{\infty})$. These are the reasons why we prefer to work with an arbitrary small interval of positives energies and
we use the approach given in  ([8], [9], [10]) which is a time-independent version of [2].

\vspace{0,3cm}\parn
So, our goal in this paper is to recover all functions $V_j$ from the knowledge of the scattering operator $S(h)$ localized near a fixed energy $\lambda$, up to  $O(h^{\infty})$
in the sense of the operator norm in $L^2 (\r^n)$.  Since we do not work with a fixed energy, but in an averaged sense,
we emphasize that the non-trapping condition is not necessary.

\vspace{0,3cm}\parn
In order to localize the scattering operator near a fixed energy
$\lambda >0$, we introduce a cut-off function $\chi \in
C_0^{\infty} (]0, +\infty [), \ \chi =1$ in a neighborhood of
$\lambda >0$.

\vspace{0,5cm}
\parn
We show that we obtain the complete asymptotic expansion
of the potential from  the knowledge of $S(h) \chi (H_0 (h))$ up to $O\ (h^{\infty})$ in ${\cal B} (L^2(\r^n ))$, when
$h \rightarrow 0$. More precisely, we determine, for $n \geq 3$,  the asymptotics expansion of the potential at infinity,
by studying the asymptotics of :
$$
F(h) = < S(h) \chi (H_0 (h)) \Phi_{h, \omega} , \Psi_{h, \omega} >
\ , \num (1.11)
$$
where $<,>$ is the usual scalar product in $L^2 (\r^n )$, and $
\Phi_{h, \omega}, \ \Psi_{h, \omega}$ are suitable test functions, (see section $2$). We need
$n\geq 3$ in order to use the uniqueness for the Radon transform in hyperplanes.


\vspace{1 cm}
\section{\bf{Semiclassical asymptotics for the localized scattering operator.}}
\par
\subsection{Definition of the test functions.}
First, let us define the unitary dilation operator $U(h^{\delta}), \
\delta >0$, on $L^2 (\r^n )$ by :
$$
U(h^{\delta}) \ \Phi (x) = h^{{ {n\delta} \over 2}} \ \Phi
(h^{\delta} x). \num (2.1)
$$
We also need an energy cut-off $\chi_0 \in C_0^{\infty} (\r^n )$
such that $\chi_0 (\xi ) =1$ if $\mid \xi \mid \leq 1$,   $\chi_0
(\xi ) =0$ if $\mid \xi \mid \geq 2$.

\vspace{0,2 cm} \parn For $\omega \in S^{n-1}$, we write $x \in
\r^n$ as $x=y +t\omega$, $y \in \Pi_{\omega}$ = orthogonal
hyperplane to $\omega$ and we consider :
$$
X_{\omega} = \{  x =y+t\omega \in \r^n \ :\ \mid y \mid \geq 1 \
\}. \num (2.2)
$$
For ${\displaystyle{\delta > {1 \over {\rho_1 -1}}}}$ and $\epsilon < 1+\delta$, we define :
$$
\fta = \ex \ U(h^{\delta}) \ \co \ \Phi  \ , \num (2.3)
$$
where  $\Phi \in  C_0^{\infty} (X_{\omega} )$ and $D = -i \nabla$, ( $\Psi_{h, \omega}$ is defined in the
same way with $\Psi \in C_0^{\infty} (X_{\omega})$).

\vspace{0,5cm}\parn

\subsection{Semiclassical asymptotics for the scattering operator.}
\vspace{0,5cm}\parn In this paper, we prove the following theorem
:

\vspace{0,3cm}\parn

\begin{th}
\hfil\break
Let $V$ be a potential satisfying $(H_2)$. Then, there exists an increasing positive sequence
${\displaystyle{(\nu_k) _{k\geq 1}}}$, (depending only on
$\delta$), with $\nu_1= \delta (\rho_1 -1)-1$ and
${\displaystyle{\lim_{k \rightarrow +\infty} \nu_k  = +\infty}}$
such that :
$$
< (S(h) -1)  \ \chi (H_0 (h)) \ \fta \ , \ \ftb >   \ \simeq \
\sum_{k=1}^{\infty} h^{\nu_k} \ < \Phi \ , \ A_k (x,\omega, D) \
\Psi> \ ,
\num (2.4)
$$
when $h \rightarrow 0$, where $A_j (x,\omega, D)$ are
differential operators defined in a recursive way.
\vspace{0,2cm} \parn
Moreover, $\forall j \geq 1, \exists \ k_j \geq 1$, (with $k_1=1$), such that :
$$
A_{k_j} (x, \omega, D) = {i \over {2 \sqrt{\lambda} }} \
\int_{-\infty}^{+\infty} \ V_j (x +t\omega) \ dt \ + \ B_j (x,
\omega, D) ,
\num (2.5)
$$
with $B_1 =0$ and for $j \geq 2, \ B_j (x, \omega, D)$ is a
differential operator only depending on the functions $V_k$,
$1\leq k \leq j-1$.
\end{th}

\subsection{Proof of Theorem 1.}
\vspace{0,5cm}\parn {\bf {Step 1 :}}

\vspace{0,3cm}\parn
Let us begin by an elementary lemma.

\begin{lem}
\hfil\break
$\forall h \ll 1$ small enough, we have :
$$
\chi (H_0 (h)) \fta \ = \ \fta .
\num (2.6)
$$
\end{lem}

\vspace{0,5 cm} \parn {\bf{Proof}}
\parn
We easily obtain :
$$
{\cal F} \ [\chi (H_0 (h)) \fta ] (\xi ) \ = \ h^{- {{n\delta}
\over 2}} \chi ((h\xi )^2 ) \chi_0 (h^{\epsilon-\delta-1} (h\xi -
{\sqrt{\lambda}}\omega )) {\cal F} \Phi (h^{- \delta -1} (h\xi -
{\sqrt{\lambda}}\omega )), \num (2.7)
$$
where ${\cal F}$ is the usual Fourier transform. Then, on $Supp \
\chi_0$, we have  $ \mid h\xi  - {\sqrt{\lambda}}\omega \mid \leq
2 h^{1+\delta - \epsilon}$. Since  $\epsilon < 1 + \delta$, we have for $h$ small enough, $\chi ((h\xi )^2 ) =1$.
$\Box$

\vspace{1 cm} \parn Then, by Lemma 2, we obtain,
$$
F(h) = < W^- (h) \fta \ , \  W^+ (h) \ftb > ,
\num (2.8)
$$
and an easy calculation gives :
$$
F(h) = < \Omega^- (h, \omega ) \co \Phi \ , \  \Omega^+ (h, \omega ) \co \Psi >, \num (2.9)
$$
where
$$
 \Omega^{\pm} (h, \omega ) = s-\lim_{ t \rightarrow \pm \infty} \
e^{it H(h, \omega )} e^{-it H_0 (h, \omega )} , \num (2.10)
$$
with
$$
H_0 (h, \omega ) = (D+  {\sqrt{\lambda}} h^{-(1 +\delta )}
\omega)^2 ,
\num (2.11)
$$
and
$$
H (h, \omega ) = H_0 (h, \omega ) + h^{-2(1+\delta )}
V(h^{-\delta} x) . \num (2.12)
$$
So, by $(2.9)$, we  have to find the asymptotics of $\Omega^{\pm} (h, \omega ) \co \Phi$.
We follow the same strategy as in [10], and we only treate only the case (+).

\vspace{0,5 cm}
\parn {\bf {Step 2 :}}
\vspace{0,3cm}\parn As in
[10], we construct, for  a suitable sequence $(\nu_k)$  defined below,
a modifier $J^+ (h, \omega )$ as a
pseudodifferential operator, (actually a differential operator close to Isozaki-Kitada's construction [4]),
having the asymptotic expansion  :
$$
J^+ (h, \omega ) = \ op \left(  1+ \sum_{k \geq 1} \ h ^{\nu_k} \
d_k^+ (x,\xi, \omega ) \right) . \num (2.13)
$$

\vspace{0,3 cm} \parn
We denote :
$$
T^+ (h, \omega ) = H(h, \omega ) J^+ (h, \omega ) - J^+ (h, \omega
) H_0 (h, \omega ) .
\num (2.14)
$$
\vspace{0,1cm}\parn
A direct calculation shows that the symbol of $T^+(h, \omega)$ is
given by :
$$
T^+ (x, \xi, h, \omega ) = h^{-1-\delta} \ \left( - \sum_{k \geq
1}  \ 2i {\sqrt{\lambda}} \omega \cdot \nabla d_k \  h^{\nu_k} +
(2i\xi \cdot \nabla d_k + \Delta d_k ) \  h^{\nu_k+1+\delta}
\right.
\num (2.15)
$$
$$
\hspace{5,7 cm}\left. +  \sum_{k \geq 1} \ V_k \
h^{-(1+\delta)+\delta\rho_k} \  + \sum_{j, k \geq 1} \ V_j \ d_k \
h^{-(1+\delta)+\delta\rho_j + \nu_k}  \right) .
$$
\vspace{0,2cm}\parn
Roughly speaking, we construct $\nu_k$ and $d_k^+$ in order to
obtain $T^+(h, \omega) = O(h^{\infty})$.
\vspace{0,1cm}\parn
First,  we choose $\nu_1 =\delta(\rho_1 -1)-1$ and we solve the
transport equation :
$$
\omega . \nabla d_1^+ (x, \omega    ) = {1 \over {2i
{\sqrt{\lambda}}}} V_1(x) . \num (2.16)
$$
The solution of $(2.16)$ is given by :
$$
d_1^+ (x, \omega ) =  {i \over {2 {\sqrt{\lambda}}}} \
\int_0^{+\infty} \ V_1 (x+t\omega ) \ dt . \num (2.17)
$$
Then, we choose $\nu_2 = min \ ( \nu_1+1+\delta, \
-(1+\delta)+\delta\rho_1, \ 2\nu_1)$ and we solve the
corresponding transport equation in order to construct $d_2$. The
functions $d_k$ and the coefficients $\nu_k$  for $k  \geq 3$ are determined in a recursive  way.
$\Box$

\vspace{0,5cm}
\parn
As in [10], we have the following proposition :

\begin{prop}
\hfil\break
$$
\Omega^+ (h, \omega ) \co \ \Phi = \lim_{t \rightarrow + \infty}
e^{it H(h, \omega )}\ J^+ (h, \omega )\  e^{-it H_0 (h, \omega )}
\co \ \Phi . \num (2.18)
$$
$$
\mid \mid (\Omega^+ (h, \omega )- J^+ (h, \omega )) \ \co \ \Phi
\mid \mid = \ O \ (h^{\infty}). \num (2.19)
 $$
\end{prop}

\vspace{1 cm}
\parn
Then, Theorem 1 follows from Proposition 3 in the same way as in [10].

\newpage

\section{\bf{The inverse scattering problem.}}

\parn
Let $V_j , \ j=1,2$,  be two potentials satisfying $(H_2)$ and let  $S_j (h)$, $j=1,2$  be the scattering operator
associated with the pair $(H_0 (h) + V_j , H_0 (h) )$.

\vspace{0,5cm}\parn
We have the
following result :

\vspace{0,3 cm}\parn

\begin{cor}
\hfil\break
For $n \geq 3$,  assume that in ${\cal B} (L^2 (\r^n ))$,
$$
\ S_1 (h) \chi (H_0 (h)) = S_2 (h) \chi (H_0 (h)) +O \ (h^\infty)\ , \ h \rightarrow 0.
\num (3.1)
$$
\parn
Then : $ V_1-V_2 \in {\cal S}  (\r^n)$ .
\end{cor}

\vspace{0,5cm}\parn
{\bf{Proof :}}
\parn
Corollary 4 follows from Theorem 1, the uniqueness for the Radon transform in hyperplanes [3],
 and the fact that $\mid \mid \fta \mid \mid$, (resp. $ \mid \mid \ftb \mid \mid$) are uniformly bounded with
respect to $h$.  We refer to [10] for the details. $\Box$

\vspace{0,8cm}\parn
{\bf{Comments.}}
\parn
It is not difficult to generalize the previous results to the case of long-range potentials,
i.e for potentials satisfying $(H_2)$ with $\rho_1>0$, using modified wave operators, close to Isozaki-Kitada's ones
[4], (see [10], Theorem 8, for details).

\vspace{0,8cm}\parn
{\bf{Acknowledgment.}}
\parn
The author is grateful to Xue-Ping Wang for suggesting this subject and fruitful remarks.
\vspace{5mm}
\parn


\begin{thebibliography}{XXXXX}


\bibitem[1]{a} I. Alexandrova, ``Structure of the Semi-Classical Amplitude for General Scattering
Relations'', Comm. in P.D.E, , vol. 30, Issues 10 to 12, p. 1505-1535, (2005).
\bibitem[2]{b} V. Enss - R. Weder, `''The geometrical approach to multidimensional
inverse scattering''', J. Math. Phys. 36 (8), p. 3902-3921, (1995).
\bibitem[3]{c} S. Helgason, ``The Radon Transform'', Progress in
Mathematics 5, Birkh\" auser, (1980).
\bibitem[4]{d} H. Isozaki - H. Kitada, ``Modified wave operators with time-independent
modifiers'', Papers of the College of Arts and Sciences, Tokyo
Univ., Vol. 32, p. 81-107, (1985).
\bibitem[5]{e} M. S. Joshi - A.S. Barreto, ``Recovering asymptotics of short range potentials'',
Communications in Mathematical Physics, 193, p. 197-208, (1998).
\bibitem[6]{f} L. Michel: ``Semi-classical limit of the scattering amplitude for trapping perturbations'',
Asymptotic Analysis , 32, p. 221-255, (2003).
\bibitem[7]{g} L. Michel: "Semi-classical behavior of the scattering amplitude for trapping perturbations
at fixed energy", Can. J. Math., 56, p. 794-824, (2004).
\bibitem[8]{h} F. Nicoleau, `''A stationary approach to inverse
scattering for Schr\" odinger operators with first order
perturbation'', Comm. in P.D.E, Vol 22 (3-4), p.  527-553, (1997).
\bibitem[9]{i} F. Nicoleau, ``An inverse scattering problem with the Aharonov-Bohm effect'',
Journal of Mathematical Physics, Issue 8, p. 5223-5237, (2000).
\bibitem[10]{j} F. Nicoleau, ``A constructive procedure to recover asymptotics of short-range or
long-range potentials'', Journal in Differential Equations 205, p. 354-364, (2004),
\bibitem[11] {ja} R. Novikov, ``${\overline{\partial}}$-method with non-zero background potential application
to inverse scattering for the two-dimensional acoustic equation'', Comm. in P.D.E, Vol 21,
(3-4), p. 597-618, (1996).
\bibitem [12] {k} D. Robert - H. Tamura, ``Semi-classical estimates for resolvents and asymptotics for
tottal scattering cross-sections'', Ann. Inst. Henri Poincar\'e, tome 46, (1), p. 415-442, (1987).
\bibitem [13] {ka} D. Robert - H. Tamura, ``Asymptotic behaviour of scattering amplitudes in
semiclassical and low energy limits'', Annales de l'institut Fourier, tome 39, (1), p. 155-192, (1989).
\bibitem [14] {n} X. P. Wang, `` Time-decay of scattering solutions and classical trajectories'',
Ann. Inst. Henri Poincar\'e Phys. Th\'eor.  47,  no. 1, 25--37, (1987).
\bibitem [15] {l} R. Weder - D. Yafaev, ``On inverse scattering at a fixed energy for potentials with a
regular behaviour at infinity'', Inverse Problems 21, p. 1937-1952, (2005).
\bibitem [16] {m} K. Yajima, "The quasi-classical limit of scattering amplitude : $L^2$ approach for
short-range potentials'', Japan. J. Math., 13, (1), p. 77-126, (1987).



\end{thebibliography}
\end{document}